# Rutherford, Radioactivity, and the Atomic Nucleus

Helge Kragh[*]

**Abstract**  Modern atomic and nuclear physics took its start in the early part of the twentieth century, to a large extent based upon experimental investigations of radioactive phenomena. Foremost among the pioneers of the new kind of physics was Ernest Rutherford, who made fundamental contributions to the structure of matter for more than three decades and, in addition, founded important research schools in Manchester and Cambridge. This paper reviews the most important aspects of Rutherford's scientific work in the period from about 1900 to 1920, and it also refers to some of his last experiments of the 1930s. The emphasis is on his theory of radioactive disintegration (1902), the discovery of the atomic nucleus (1911), and the first artificially produced element transformation (1919). Following the transmutation experiments, Rutherford developed elaborate models of the atomic nucleus, but these turned out to be unsuccessful. Other subjects could be included, but the three mentioned are undoubtedly those of the greatest importance, the nuclear atom perhaps the greatest and the one with the most far-reaching consequences.

## 1. The career of a young physicist

One of the most eminent physicists ever, Ernest Rutherford earned his scientific reputation primarily by his pioneering contributions to radioactivity and nuclear physics (Eve 1939; Wilson 1983). Indeed, he discovered the atomic nucleus and founded nuclear physics as a flourishing research field. When one of his colleagues and biographers, the English-Canadian physicist Arthur Stewart Eve, once observed that Rutherford always appeared to be riding on the "crest of the wave," Rutherford replied, "Well! I made the wave, didn't I?" (Eve 1939, p. 436). He was

---

[*] Centre for Science Studies, Aarhus University, Aarhus, Denmark. E-mail: helge.kragh@ivs.au.dk. A version of this paper will appear in Spanish in a book to be published by the Residencia de Estudiantes in Madrid.

not a modest man, and had no reason to be modest, yet he added, "At least to some extent."

As a result of his early work in radioactivity, at the age of 37 he was awarded the Nobel Prize in chemistry. That he was awarded the chemistry prize, and not the one in physics, may appear strange, but at the time radioactivity was generally considered a branch of chemistry rather than physics. Later in life, when he had become a public figure and statesman of science, he received numerous honours and scientific prizes. For example, in 1914 he was made a knight, in 1922 he received the prestigious Copley Medal from the Royal Society, and in 1925 he was honoured by being conferred the Order of Merit. He ended his life as Baron Rutherford of Nelson, a title which was conferred on him in 1931.

Born in 1871 near Nelson in rural New Zealand, in 1887 young Rutherford won a scholarship to Nelson College, a secondary school, where he boarded for three years. After having received another scholarship, from 1890 to 1894 he attended Canterbury College in Christchurch, where he was able to cultivate his growing interest in mathematics and physics. Having obtained a Master of Arts degree, he did work on the magnetization of iron by a rapidly alternating electric current and also invented an apparatus that could detect wireless or Hertzian waves over what at the standards of the time were large distances. In 1895 he was awarded a scholarship that allowed him to go to England to do graduate research at Cambridge University's famous Cavendish Laboratory. The director of the laboratory was J. J. Thomson, who a few years later would initiate a new chapter in the history of physics by discovering the electron. Thomson was keenly interested in Rutherford's electromagnetic detector and generally impressed by his brilliance at experimental research.

The result was a joint paper on the ionization of gases caused by the recently discovered X-rays. According to the two Cambridge physicists, the effect of the X-rays was to create an equal number of positive and negative ions in the gas molecules. Apart from examining the action of the X-rays, in subsequent work Rutherford (1899) turned to the even more recent – and even more enigmatic – rays produced by uranium and a few other substances. In his first work on what at the time was often known as either "uranium rays" or "Becquerel rays" (after their discoverer, Henri Becquerel) he demonstrated that the rays were inhomogeneous, consisting of at least two components. One of the components was easily absorbed while the other had a greater ability of penetrating matter. Rutherford called the first component "alpha" and the second type "beta." At the end of the lengthy paper he suggested that the rays were similar to X-rays, but "The cause and origin of the radiation continuously emitted by uranium and its salts still remain a mystery" (Rutherford 1899, p. 163).

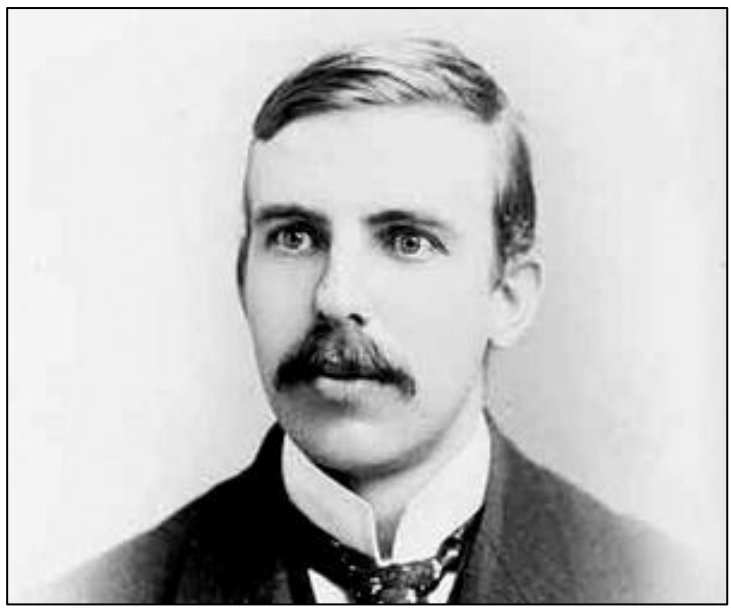

Figure 1. Young Ernest Rutherford, at the time he was a professor at McGill University.

Shortly after having completed this important work on uranium rays, Rutherford left Cambridge for Montreal, Canada, where he had been offered the MacDonald professorship in physics at McGill University (Figure 1). It was here he would make a name for himself as the leading authority in radioactive research.

## 2. Enigmas of radioactivity

Following the announcement of Marie and Pierre Curie in 1898 that they had discovered two new radioactive elements, radium and polonium, the nascent field of radioactivity was eagerly taken up by an increasing number of physicists and chemists. Which substances were radioactive? How did they fit into the periodic system of the elements? Was the apparently spontaneous activity affected by physical and chemical changes? These were some of the questions addressed by the first generation of specialists in radioactive research, a field initially dominated by scientists in Paris and Montreal. The ambitious and competitive Rutherford was determined to establish himself as the leader of research in radioactivity. "I have to keep going, as there are always people on my track," he wrote to his mother in 1902. "I have to publish my present work as rapidly as possible in order to keep in the race. The best sprinters in this road of investigation are Becquerel and the Curies in Paris, who have done a great deal of very important work on the subject of radioactive bodies during the last few years" (Pais 1986, p. 62).

Rutherford not only kept in the race, his scientific results established him as an authority in radioactivity on line with, or even in front of, the Curies. During his Montreal period he engaged not only in innovative research but also found time to write two of the first comprehensive textbooks on the subject: *Radio-Activity* appeared in 1904, and two years later his Silliman lectures of 1905 were published under the title *Radioactive Transformations*. They are both classics of science. Examining the ionization caused by thorium, in 1900 he identified a radioactive gas that became known as "thorium emanation" but the chemical nature of which at first eluded him. The puzzle only increased when somewhat similar "emanations" were found associated with radium and actinium. Collaborating with Frederick Soddy, an English demonstrator in chemistry working at McGill, Rutherford provided experimental evidence that the emanation was most likely a new inert gas – what today is called radon. Moreover, they found that emanation, whether from thorium or radium, lost its activity over a brief period of time. This was the first indication that radioactive substances were characterized by a new parameter, a life-time or a half-life (Rutherford 1900).

More than other specialists in radioactivity, Rutherford focused his research on the properties and nature of the alpha rays. Early experiments suggested that these rays went undeflected through electric and magnetic fields and thus were neutral, a view that Rutherford and other researchers held for a year or two. However, further experiments made in Montreal proved that the particles making up the alpha rays were actually positive and with a mass comparable to that of the hydrogen atom. Rutherford soon came to the conclusion that the particles were positive ions of helium, the inert gas that originally had been detected in the sun's atmosphere and in 1895 was discovered in uranium minerals. Although convinced that alpha particles were doubly charged helium ions ($He^{2+}$), it was only in 1908, after he had moved to Manchester, that he provided the final proof for the $\alpha = He^{2+}$ hypothesis. This he did in a brilliant experiment with his assistant Thomas Royds in which it was proved spectroscopically that helium was produced by the alpha rays emitted by radium emanation. Together with data from the magnetic and electric deflection of alpha rays, the experiment gave "a decisive proof that the $\alpha$ particle after loosing its charge is an atom of helium" (Rutherford and Royds 1909, p. 166).

While primarily investigating radioactivity from an experimental perspective, Rutherford was also keenly interested in explaining the phenomenon in causal terms. The origin of radioactivity was at the time a complete mystery, but during the first decade of the twentieth century many physicists came to believe that it had to be found in the internal structure of the atom (Kragh 1997).

Rutherford was generally in favour of a model of the composite atom along the lines suggested by J. J. Thomson, assuming that negative electrons and positive alpha particles preexisted in the atom in some dynamical equilibrium configuration. As they rotated or vibrated, they would loose energy in accordance with the laws of electromagnetism and eventually be expelled from the atom.

Rutherford realized that this "radiation-drain hypothesis" was not entirely satisfactory, but in lack of better explanation he cautiously supported it. As he saw it, some future development of atomic theory, perhaps a modification of the Thomson model, would most likely yield a causal explanation of radioactivity in terms of the internal structure of the atom. He examined the question in some detail in his Silliman lectures, where he concluded that Thomson's approach, although "somewhat artificial," was nonetheless "of great value as indicating the general method of attack of the greatest problem that at present confronts the physicist." Rutherford (1906, p. 265) wrote:

> As our knowledge of atomic properties increases in accuracy it may yet be possible to deduce a structure of the atom which fulfills the conditions required by experiment. A promising beginning has already been made, and there is every hope that still further advances will soon be made in the elucidation of the mystery of atomic structure.

The mystery of atomic structure was eventually elucidated, but an explanation of radioactivity of the kind Rutherford had in mind never appeared. Only after the emergence of quantum mechanics did it become clear that radioactivity is a genuinely acausal phenomenon. In any case, on the hypothesis that radioactivity was caused by disruptions of electrical particles in the interior of the atom it seemed natural to suppose that it was a property common to all matter. Were all elements radioactive, only some more active than others? This was indeed what many scientists believed at the time, especially after Norman Campbell had proved in 1906 that potassium and rubidium are weakly radioactive (they contain the beta-radioactive isotopes K-40 and Rb-87, respectively, both with very long life-times).

Rutherford shared the view that radioactivity is a general property of matter: "It is a matter of general experience that every physical property discovered for one element has been found to be shared by others in varying degree … It might thus be expected as general principles that the property of radioactivity which is so marked in a substance like radium would be shown by other substances" (Rutherford 1906, p. 217).

## 3. Transmutations and atomic decay

In December 1908 Rutherford received in Stockholm the Nobel Prize in chemistry "for his investigations into the disintegration of the elements, and the chemistry of radioactive substances." Somewhat bemused to have transformed so quickly from a physicist to a chemist, he chose as the subject for his Nobel Lecture "The Chemical Nature of the Alpha Particles from Radioactive Substances" (Rutherford 1962-1965, vol. 2, pp. 137-146). Rutherford had been nominated for the 1907 prize in both physics (seven nominations) and chemistry (one nomination), and also in 1908 there were more physics than chemistry nominations, namely, five to three. Interestingly, far most of the nominators were Germans, while none were British (Jarlskog 2008). An important reason for the award of the prestigious prize was the disintegration or decay theory of radioactivity, which he had proposed some years earlier and which still counts as the fundamental law of radioactive change.

As mentioned, in his study of 1900 of the properties of thorium emanation Rutherford had found that the activity of the substance decreased exponentially in time and thus could be ascribed a definite half-life. In his further investigation of the phenomena of decay and regeneration of radioactive intensity he joined forces with Soddy, which in 1902 resulted in the first version of the disintegration theory, to appear in a refined form the following year (Figure 2). According to Rutherford and Soddy, a radioactive substance transformed into another substance in the sense that the atoms changed from a "parent element" to another "daughter element" at a characteristic rate. Not only do atoms change or transmute, they also do so randomly, such as expressed by a certain decay constant ($\lambda$) that depends only on the nature of the radioactive element. The fraction of atoms $dN$ that disintegrate in a small interval of time $dt$ follows the expression

$$dN = -\lambda N(t)dt$$

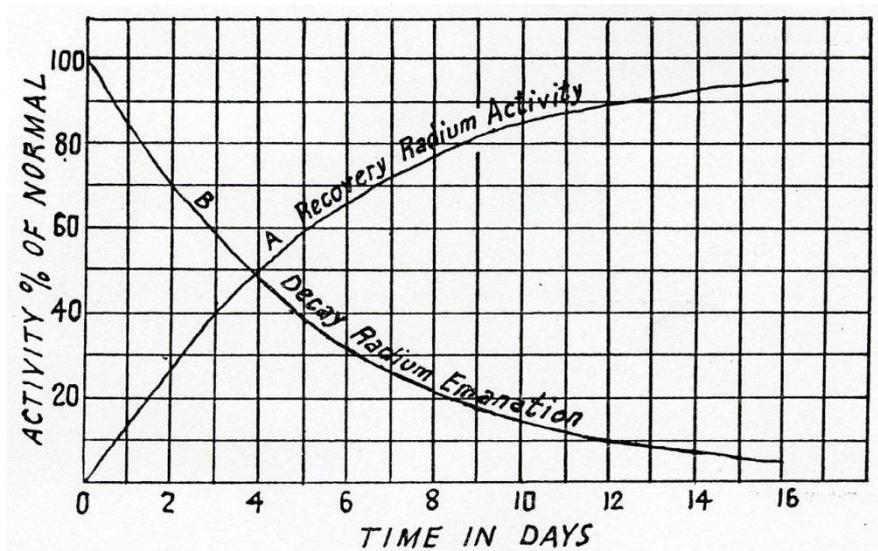

Figure 2. The exponential decay of radium emanation (radon) and the regeneration of radium, as shown in the 1903 Rutherford-Soddy paper.

In another formulation, if a radioactive substance originally, at $t$ = 0, consists of $N_0$ atoms, after a time $t$ the number of the same atoms will be reduced to

$$N(t) = N_0 e^{-\lambda t}$$

The decay constant, which is measured in inverse time units, can also be expressed by the mean life-time $T$ or the half-life $T_{1/2}$, the time required for the number of particles to decrease by a factor of 2. The connection between the three quantities is

$$\lambda = \frac{1}{T} = \frac{ln2}{T_{1/2}} = \frac{0.693}{T_{1/2}}$$

As Rutherford was well aware, the decay law is of a statistical nature, giving only the probability that an atom decays in some interval of time between $t_0$ and $t_0 + \Delta t$. Some atoms will decay almost instantly, while others will survive for a much longer time. Another way of expressing this statistical nature is that a radioactive atom does not age: the probability of decay does not depend at all on the age of the atom, but only on its kind. Rutherford realized that the form of the decay law seemed to disagree with causal-dynamical models of radioactivity, such as advocated by J. J. Thomson. For according to these models, "All atoms formed at the same time should last for a definite interval. This, however, is contrary to the observed law of transformation, in which the atoms have a life embracing all values from zero to infinity" (Rutherford 1906, p. 267). Neither Rutherford nor others could explain the conundrum.

Although the transformation theory of Rutherford and Soddy met some resistance in chemical circles, it was accepted remarkably smoothly by the majority of physicists. However, French physicists were an exception. Marie and Pierre Curie, Albert Laborde, André Debierne, and other French specialists in radioactivity adopted a positivistic ideal of science which implied that phenomena should be given high priority and material hypotheses low priority. They consequently refused to commit themselves to specific causal hypotheses such as subatomic transformation (Malley 1979). The resistance in Paris towards the Rutherford-Soddy theory was a major reason why the momentum in radioactive research changed from Paris to other centres in Canada, England and Germany.

Rutherford and Soddy realized that the constant activities of uranium and thorium were only apparent, a result of very long life-times that would not immediately turn up experimentally. In an important paper of 1903 they stated the general law of radioactive change, noting that "The complexity of the phenomena of radioactivity is due to the existence as a general rule of several different types of matter changing at the same time into one another, each type possessing a different radioactive constant" (Rutherford and Soddy 1903, p. 580). They also pointed out, as did Pierre Curie and Laborde in Paris at the same time, that the energy released in the disintegration of radium was enormous compared with the one arising from chemical combustion processes. For the energy of one gram of radium they estimated a lower limit of 15,000 calories or 63 kJ per year.

Rutherford and Soddy were evidently fascinated by the prospects of what they called "atomic energy" and which they conceived as a general form of energy locked up in all atoms, not only in those belonging to the radioactive elements. They further speculated that the new atomic energy would prove to be of great importance to astrophysics as it promised an explanation of stellar energy production: "The maintenance of solar energy, for example, no longer presents any fundamental difficulty if the internal energy of the component elements is considered to be available, i.e. if processes of sub-atomic change are going on."

Less speculatively, Rutherford and his American collaborator, the radiochemist Bertram Boltwood at Yale University, realized that the radioactive decay law might provide a way to estimate the age of old rocks and thus to settle the much discussed question of the age of the earth, a question that physicists and geologists had widely different opinions of (Badash 1989). Recognizing that helium was produced by alpha-radioactive bodies, in 1904 Rutherford suggested that the helium trapped in radioactive minerals might provide a means for determining the age of the earth. The method did not work, and a better one was proposed by Boltwood, who had found that lead is always present in uranium minerals. This he took as evidence that it was the end product of the uranium series. Using Rutherford's data and the decay law, the age of rock samples could be estimated. The result was that the age of the earth was probably greater than one billion years, a value much exceeding earlier estimates based on thermodynamical reasoning.

In 1929 Rutherford used a much improved version of the uranium-lead method to argue that the upper limit of the age of the earth was 3.4 billion years, a value some two billion years short of the presently known age, which is $4.55 \pm 0.07$ billion years (Rutherford 1929).

## 4. The road to the nuclear atom

Rutherford was not originally much interested in atomic structure, except that he in a general way favoured a model of the kind proposed by Thomson, where the electrons moved in circular orbits within a sphere of atomic dimensions homogeneously filled with a fluid of positive electricity. "The mobile electrons constitute, so to speak, the bricks of the atomic structure, while the positive electricity acts as the necessary mortar to bind them together," as Rutherford (1906, p. 266) phrased it. When he seriously turned to atomic theory, which he only did after having become professor at Manchester University in 1908, it was to a large extent the result of his deep interest in the particles constituting the alpha rays (Heilbron 1968; Badash 1983).

In experiments with Hans Geiger, a young German physicist, Rutherford developed a technique that enabled them to detect individual alpha particles by the scintillations caused when a particle hit a screen of zinc sulphide. In 1908 Geiger reported preliminary results of the scattering of alpha particles on metal foils, but only at small angular deflections. The following year he investigated the matter more fully in collaboration with 21-year-old Ernest Marsden (Figure 3). The two physicists found that heavier metals were far more effective as reflectors than light ones and that a thin platinum foil scattered one of every 8,000 alpha particles by an angle φ larger than 90 degrees. These early experiments induced Rutherford to develop a unified scattering theory that could not only account for the ordinary scattering of alpha and beta particles but also for the large-angle scattering of the first kind of particle.

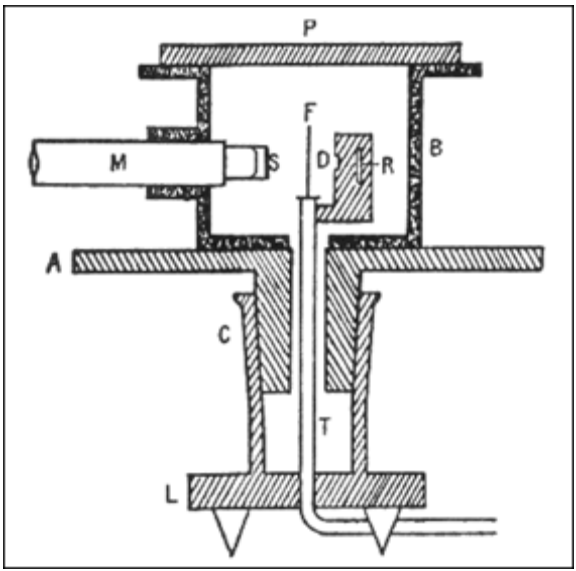

Figure 3. The Geiger-Marsden apparatus for the detection of the angular distribution of scattered alpha particles. The microscope M is provided with a screen S of zinc sulphide. B and M can rotate, while the metal foil F and the radon source R remain in their positions.

Whereas Thomson pictured the alpha particle as an atomic congregation of 10-12 electrons, Rutherford came to the conclusion that it was a point particle, like the electron. Because the alpha particle was a helium atom deprived of its two electrons, this view implied, in effect, a nuclear model of the helium atom. By late 1910 Rutherford was focusing on a new picture of atomic structure that was consistent with the scattering experiments and differed drastically from the Thomson model (Figure 4). On 14 December 1910 he wrote to Boltwood: "I think I can devise an atom much superior to J. J.'s, for the explanation of and stoppage of $\alpha$ and $\beta$ particles, and at the same time I think it will fit in extraordinarily well with the experimental numbers" (Badash 1969, p. 235). He presented his new and supposedly superior atomic model in a landmark paper in the *Philosophical Magazine* of May 1911.

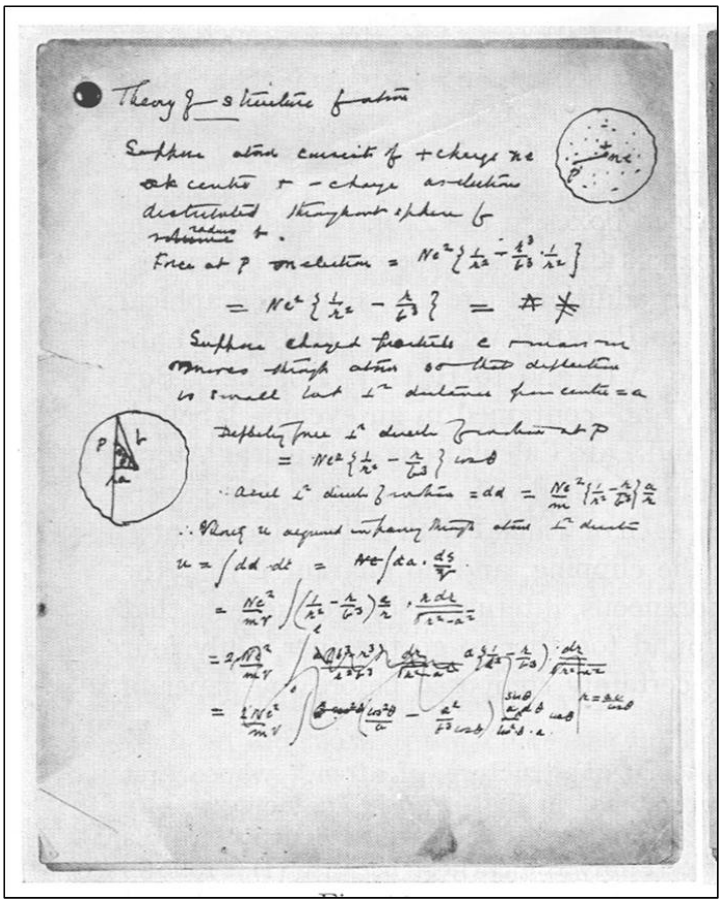

Figure 4. Rutherford's early calculations of the scattering of alpha particles, and his sketch of the nuclear atom. He pictured a heavy atom as consisting of a central positive charge surrounded by negative electrons, but without considering their configurations.

In this paper Rutherford concluded that in order to produce the observed deflections of $\phi > 90°$, scattering had to take place in a single encounter between the alpha particle and a highly charged and concentrated mass. He consequently suggested that the atom contained at its centre a massive charge $Ne$ surrounded by a cloud of the opposite charge (*e* denotes the elementary electrical charge). Since the results of his calculations were independent of the sign of the charge, the nucleus could just as well be a concentration of electrons embedded in a positive fluid – not unlike an extreme case of the Thomson atom. He explained: "Consider an atom which contains a charge ± *Ne* at its centre surrounded by a sphere of electrification containing a charge ∓ *Ne* supposed uniformly distributed throughout a sphere of radius *R*." As to the sign of the central charge: "For convenience, the sign will be assumed to be positive" (Rutherford 1911, p. 670 and p. 687).

Based on this picture of the atom, Rutherford derived a general formula that expressed the number of charged particles *y* scattered a certain angle ϕ at a distance from the scattering material. The formula, soon known as the Rutherford scattering formula, related the function $y = y(\phi)$ to the mass and velocity of the incident particles, the number of atoms in a unit volume of the scattering material, and the nuclear charge *N* of this material. As Rutherford demonstrated, in the case of alpha particles in particular his formula agreed very well with the experimental data obtained in Manchester. The data indicated "that the value of this central charge for different atoms is approximately proportional to their atomic weights, at any rate for atoms heavier than aluminium." According to Rutherford's analysis, the gold atom, of atomic weight A = 197, had a charge of $N \cong 100$, which agreed reasonably well with what he suspected was a general approximation, namely the relationship A/2 < *N* < A.

It should be pointed out that in 1911 the notion of the atomic number Z as the ordinal number of the elements in the periodic system did not yet exist. Chemical elements were still defined by their atomic weights. Rutherford might have been the first to use the term "atomic number," which he did at the end of 1913, but the concept was introduced a little earlier by the Dutch amateur physicist Antonius van den Broek. According to van den Broek, the physical meaning of the atomic number was that it represented the nuclear charge in the Rutherford atom (that is, Z = *N*).

The idea of concentrating the positive atomic charge in a tiny central part of the atom was not quite original with Rutherford's model. In 1904 the Japanese physicist Hantaro Nagaoka had proposed a "Saturnian" model in which thousands of electrons rotated on one or more circles around a massive positive charge of minute dimensions. Nagaoka's Saturnian atom was well known for a couple of years, but it turned out to be unstable and played no motivating role for Rutherford. He did however refer to Nagaoka's theory, which he compared with his own. Nowhere in his paper of 1911 did Rutherford refer to the central charge as a "nucleus" or his model as a "nuclear atom" (nor did Nagaoka use such terminology). The name "nucleus" was first used by the British astrophysicist John Nicholson in a paper that appeared later in 1911, and then in connection with an atomic model which was only superficially related to Rutherford's.

Terminology apart, the nuclear atom introduced by Rutherford in 1911 did not make a splash in the world of physics. It would soon be recognized as a revolution in the theory of matter, but at the time it was met with indifference and scarcely considered to be a new theory of the constitution of the atom. It was not mentioned in proceedings of the first Solvay congress, taking place in Brussels in the autumn of 1911 with Rutherford as a participant, nor did it receive much attention in the physics journals. Even Rutherford himself did not consider his discovery as the epoch-making event that it turned out to be. For example, in his massive 1913 textbook on radioactivity, titled *Radioactive Substances and their Radiations*, there were only two references to the nuclear atom and its implications. He now declared the nucleus to be positively charged, surrounded by electrons "which may be supposed to be distributed throughout a spherical volume or in concentric rings in one plane." The nucleus was extremely small, but not point-like. On the contrary, Rutherford pictured it as a complex body held together by what would become known as nuclear forces, the first example of strong interactions (Rutherford 1913, p. 620):

> Practically the whole charge and mass of the atom are concentrated at the centre, and are probably confined within a sphere of radius not greater than $10^{-12}$ cm. No

> doubt the positively charged centre of the atom is a complicated system in movement, consisting in part of charged helium and hydrogen atoms. It would appear as if the positively charged atoms of matter attract one another at very small distances for otherwise it is difficult to see how the component parts at the centre are held together.

In other words, if a strong attractive force was not postulated, the electrostatic repulsion between the constituent nuclear charges would blow the nucleus apart.

It is customary to speak of Rutherford's atomic model, but in 1911 there was not really such a model, at least not in the sense of "atomic model" that was ordinarily adopted at the time. Rutherford presented his theory primarily as a scattering theory and realized that, considered as a theory of atomic structure, it was most incomplete. First and foremost, it could offer no suggestion of how the extra-nuclear electrons were arranged, the very issue that was central to atomic models. "The question of the stability of the atom proposed need not be considered," he wrote, "for this will obviously depend upon the minute structure of the atom, and on the motion of the constituent charged parts" (Rutherford 1911, p. 671). His nuclear atom was impotent when it came to chemical questions such as valency and the periodic system, and it fared no better when it came to physical questions such as spectra, magnetic properties, and the dispersion of light. An atomic theory anno 1911 would be considered really convincing only if it included the system of electrons, which Rutherford's did not.

The status of the nuclear theory improved in the spring of 1913, when Geiger and Marsden published new data on the scattering of alpha particles that were in excellent agreement with the scattering formula. The new experiments sharpened the relationship between the nuclear charge and atomic weight, which Rutherford now took to be $N \cong A/2$. The work of Geiger and Marsden confirmed Rutherford's atomic model considered as a scattering theory, but not as a theory of atomic structure. The new results obtained in Manchester were as irrelevant for the electronic configurations as Rutherford's atom was silent about them. It is understandable that most physicists outside Manchester preferred to ignore the nuclear atom or consider it merely a hypothesis.

## 5.  The Bohr-Rutherford atomic model

After a stay with Thomson in Cambridge, in March 1912 young Niels Bohr arrived in Manchester to do postdoctoral work under Rutherford. The 26-year-old Danish physicist was convinced of the essential truth of the nuclear atom, which he first used to investigate the energy loss of alpha particles as they were absorbed in matter. Within a few months he was able to present to Rutherford some preliminary ideas of atomic structure that a year later resulted in a seminal series of three papers with the common title "On the Constitution of Atoms and Molecules" (Heilbron and Kuhn 1969; Kragh 2012). Bohr's new theory turned Rutherford's nuclear atom into a full-blown quantum model of the atom, and it was only with this extension that the nuclear atom became scientifically fertile and widely accepted. For example, Kasimir Fajans, a Polish radiochemist who worked in Manchester, did not originally accept Rutherford's nuclear model. In a letter to Rutherford from the end of 1913, he wrote: "I have followed Bohr's papers with extraordinary interest, and now I no longer doubt the complete correctness of your atomic theory. The reservations I expressed ... have been entirely removed by Bohr's work" (Kragh 2012, p. 48).

In his original discussion of the Rutherford atom Bohr focused on its lack of mechanical stability and the formation of simple molecules, whereas he disregarded issues of optical spectroscopy. Only in March 1913 did he fully realize that the new atomic theory he had in mind must include the emission of line spectra as a crucial feature. In his paper in the July 1913 issue of *Philosophical*

*Magazine* he emphasized from the beginning that all his considerations rested on Rutherford's nuclear atom, which he summarized as follows (Bohr 1913, p. 1):

> According to this theory, the atoms consist of a positively charged nucleus surrounded by a system of electrons kept together by attractive forces from the nucleus; the total negative charge of the electrons is equal to the positive charge of the nucleus. Further, the nucleus is assumed to be the seat of the essential part of the mass of the atom, and to have linear dimensions exceedingly small compared with the linear dimensions of the whole system.

Bohr pointed out several weaknesses of this model, in particular that it was unstable from both a mechanical and electromagnetic point of view. In the case of a hydrogen atom, consisting of a single electron revolving around a nucleus of opposite charge, it followed from Maxwellian electrodynamics that the electron would radiate energy, with the result that the radius would decrease until the electron coalesced with the nucleus (Figure 5). Obviously, to save the attractive idea of the nuclear atom some drastic revisions in the picture of the atom had to be made. These revisions Bohr formulated in two postulates which violated assumptions of classical physics and introduced the quantum discontinuity as an essential feature in atomic architecture. An atomic system, he claimed, can only exist in certain "stationary states" in which revolving electrons do not emit energy. Only when the system changes abruptly from a higher state $E_2$ to a lower state $E_1$ will the energy difference appear as radiation with a frequency $\nu$ given by

$$E_2 - E_1 = h\nu \,,$$

where $h$ is Planck's quantum of action. What matters in the present context is that Bohr's theory was eminently successful in explaining precisely those phenomena about which the original Rutherford atom had nothing to say. The successes were primarily restricted to the spectra of simple atoms and ions (such as H, He$^+$ and

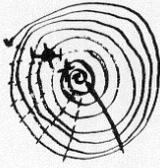

Figure 5.  Bohr's draft calculations from 1913 of the radiation emitted
by a circulating electron, causing it to spiral towards the nucleus.

Li²⁺), but the theory also promised an explanation of the formation of molecules and

the structure of complex atoms in accordance with the periodic system

Although Rutherford's tiny atomic nucleus was largely outside the scope of

Bohr's theory, it was a necessary foundation for it and sometimes appeared

explicitly in the theory. For example, Bohr was able to explain the spectroscopic

Rydberg constant $R$ in terms of fundamental constants of nature (the mass of the electron $m$, Planck's constant $h$, the elementary charge $e$, and the speed of light $c$).

When it turned out that the predicted hydrogen spectrum did not agree precisely with measurements, Bohr pointed out that the electron and the nucleus would both rotate around their common centre of mass. In this case the mass of the electron $m$ should be replaced with the so-called reduced mass

$$\frac{mM}{m + M} = \frac{m}{1 + m/M},$$

where $M$ is the mass of the nucleus. The result was that Rydberg's constant would depend slightly on the mass of the nucleus, and with this correction the disagreement vanished.

Physicists accepting the nuclear model generally assumed that beta particles came from the rings of electrons surrounding the nucleus. This was also the view of Rutherford, who distinguished between "the instability of the central nucleus and the instability of the electronic distribution. The former type of instability leads to the expulsion of an $\alpha$–particle, the latter to the appearance of $\beta$ and $\gamma$–rays" (Rutherford 1912, p. 461). However, according to Bohr it was necessary that beta particles had the same origin in the nucleus as alpha particles. It was known that some radioactive substances, apparently belonging to the same element, emitted beta rays with different velocities. If the substances were isotopes they would have the very same electron systems and only differ in their atomic weights, meaning their nuclei. As Bohr pointed out, this showed that beta rays must have their origin in the nucleus and not in the electronic system. Bohr's conclusion forced Rutherford to change his mind and it soon became generally accepted among specialists in radioactivity.

In 1914 James Chadwick, one of Rutherford's former students, had shown that the energy of the beta spectrum was essentially continuously distributed, contrary to the discrete energies of the alpha spectrum (Figure 6). This was a result that would cause a great deal of problem in physics until it was finally explained by the neutrino hypothesis in the early 1930s (Jensen 2000). Rutherford thought he could account for the continuous spectrum on the assumption that beta particles were expelled from the nucleus. In passing through the outer distribution of electrons a beta particle would collide with these electrons and share its energy with them. "As a statistical result of a large number of atoms," he wrote, "the velocity of the escaping β particles will, on the average, be continuously distributed … [and] this would give rise to the continuous spectrum of β rays" (Rutherford 1914b, p. 308). Alas, it soon turned out that Rutherford's new theory of beta radioactivity was no more able to explain the continuous spectrum than other theories.

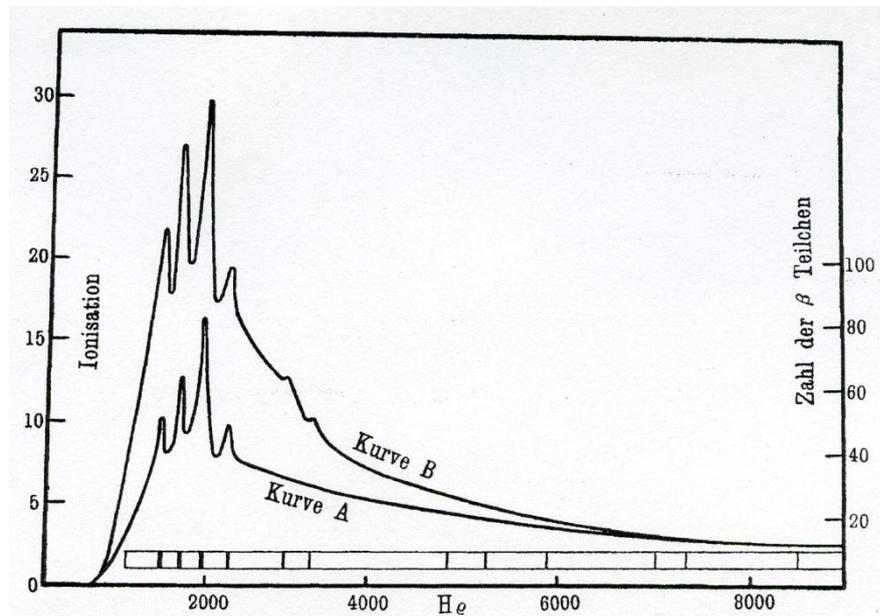

Figure 6. Chadwick's beta-ray spectrum of 1914, obtained by two different methods.

Realizing that Bohr's theory provided new confirmation of the nuclear atom, Rutherford supported it from the very beginning. However, although advocating the theory he was somewhat reluctant when it came to its foundation in the postulates and its use of quantum theory. In a paper of 1914 he expressed the feeling of many physicists: "There no doubt will be much difference of opinion as to the validity of the assumptions made by Bohr in his theory of the constitution of atoms and molecules, but a very promising beginning has been made on the attack of this most fundamental of problems, which lies at the basis of physics and chemistry" (Rutherford 1914a, p. 351). Rutherford sensed that Bohr's theory broke with established norms of physics, such as causality, and this made him uneasy. As early as March 1913, before the publication of the theory, he responded in a letter to Bohr's ideas about quantum jumps (Kragh 2012, p. 70):

> There appears to me one grave difficulty in your hypothesis, which I have no doubt you fully realize, namely, how does an electron decide what frequency it is going to vibrate at when it passes from one stationary state to the other? It seems to me that you would have to assume that the electron knows beforehand where it is going to stop.

Bohr's atomic theory was strange indeed, but it was also empirically successful and the best possible support of the nuclear atom. Another very important support came at the end of 1923, when Henry Moseley announced his first investigations of the characteristic X-rays from a large number of elements, a line of work he had begun in Manchester. Moseley's data only made sense on the assumption that a chemical element was defined by the positive charges in the nucleus; they showed that each place in the periodic table corresponded to a change of one nuclear charge. "The results," he wrote, "have an important bearing on the question of the internal structure of the atom, and strongly support the views of Rutherford and of Bohr" (Moseley 1913, p. 1025).

Latest by 1916 the nuclear atom, which now had become the Bohr-Rutherford model of the atom, was generally accepted in the physics community.

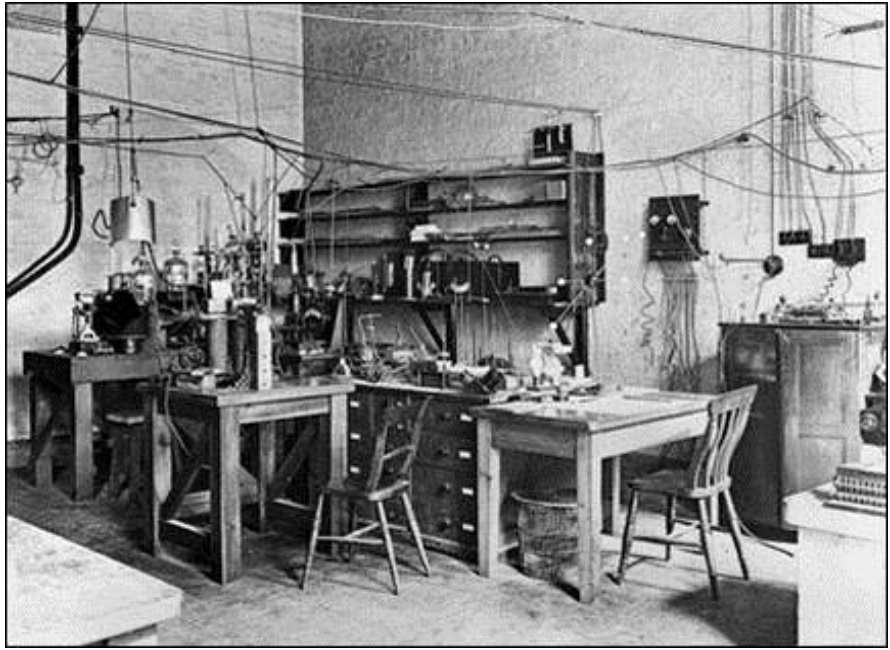

Figure 7. The Manchester laboratory in which Rutherford made the first experiments with artificial transmutations of elements.

## 6. Speculations and artificial transmutations

While Bohr continued to examine the general principles of quantum theory and atomic structure, Rutherford focused in his work after 1915 on the secrets of the atomic nucleus. In 1919 he left Manchester to succeed Thomson as director of the Cavendish Laboratory in Cambridge, but before leaving he had made another spectacular discovery, almost as important as the transformation theory of radioactivity and the nuclear atom (Badash 1983; Wroblewski 2002). In December 1917, while still engaged in some war-related work, he wrote to Bohr in Copenhagen: "I am detecting & counting the lighter atoms set in motion by $\alpha$ particles & the results, I think, throw a good deal of light on the character & distribution of forces near the nucleus. I am also trying to break up the atom by this method" (Stuewer 1986, p. 322). What he in a later letter to Bohr called "some rather startling results" were published in the April 1919 issue of the *Philosophical Magazine*.

In a study of 1914 of the action of alpha particles on a hydrogen gas Marsden had obtained results that seemed to indicate that the radioactive source itself gave off hydrogen nuclei. He suspected that the hydrogen nuclei were being emitted along with the alpha particles. In a reinvestigation of these experiments, Rutherford studied systematically the action of alpha particles on various gases by detecting the scintillations produced by particles formed by the action (Figure 7). With alpha rays from radium C (bismuth-214) passing through pure nitrogen he observed what he called an anomalous effect, namely, the production of long-range scintillations that appeared much the same as those produced by hydrogen atoms. Rutherford (1919, p. 586) found it

> … difficult to avoid the conclusion that these long-range atoms arising from the collision of alpha particles with nitrogen are not nitrogen atoms but probably charged atoms of hydrogen, or atoms of mass 2. If this be the case, we must conclude that the nitrogen atom is disintegrated under the intense forces developed in close collision with a swift $\alpha$ particle, and that the hydrogen atom which is liberated formed a constituent part of the nitrogen nucleus.

Written in a later notation, the process suggested by Rutherford was

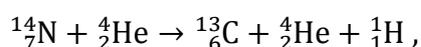

$$^{14}_{7}\text{N} + {}^{4}_{2}\text{He} \rightarrow {}^{13}_{6}\text{C} + {}^{4}_{2}\text{He} + {}^{1}_{1}\text{H} \, ,$$

an example of modern alchemy! At the end of his paper, Rutherford spelled out the consequences of his interpretation of the experiment as the first case of an artificial atomic disintegration, although one performed by projectiles of natural origin:

> Considering the enormous intensity of the forces brought into play, it is not so much a matter of surprise that the nitrogen atom should suffer disintegration as that the $\alpha$ particle itself escapes disruption into its constituents. The results as a whole suggest that, if $\alpha$ particles – or similar projectiles – of still greater energy were available for experiment, we might expect to break down the nuclear structure of many of the lighter atoms.

Further work done in the Cavendish Laboratory proved that Rutherford had indeed achieved an artificial transmutation of one element to another, but not quite the one he thought. Believing that the alpha particle was a fundamental constituent of matter, he could not imagine that it would itself be transformed. In 1925 Patrick Blackett, a junior colleague of Rutherford's and a future Nobel laureate, used the cloud chamber to photograph eight cases of alpha-nitrogen disintegration out of a total of 23,000 photographs with about 420,000 tracks of alpha particles. His photographs showed two and not three tracks emerging from the impact point, and the process was consequently reinterpreted as a nitrogen-oxygen transmutation:

$$^{14}_{7}\text{N} + {}^{4}_{2}\text{He} \rightarrow {}^{17}_{8}\text{O} + {}^{1}_{1}\text{H}$$

In a lecture of 1925 to the Royal Institution, Rutherford (1925, p. 588) explained: "He [Blackett] concluded that the $\alpha$ particle was captured in a collision which led to the ejection of a proton. … It thus appears that the nucleus may increase rather than diminish its mass as the result of collisions in which a proton is expelled."

Rutherford considered the nuclear experiments made in Manchester and Cambridge to be of particular interest because they indicated how the atomic nucleus was constituted. In agreement with the so-called "two-particle paradigm,"from 1913 to about 1933 it was generally assumed that the nucleus consisted of hydrogen nuclei (protons) and electrons. For example, nitrogen-14 was thought to consist of 14 protons and 7 electrons, and lithium-7 of 7 protons and 4 electrons. However, there were many ideas of other, more or less hypothetical constituents. In his Bakerian lecture delivered to the Royal Society on 3 June 1920, Rutherford offered several suggestions of new nuclear particles, although ultimately these would all consist of protons and electrons. Incidentally, this was the occasion at which he introduced the name "proton" as an alternative to hydrogen nucleus, hydrogen ion, or positive electron.

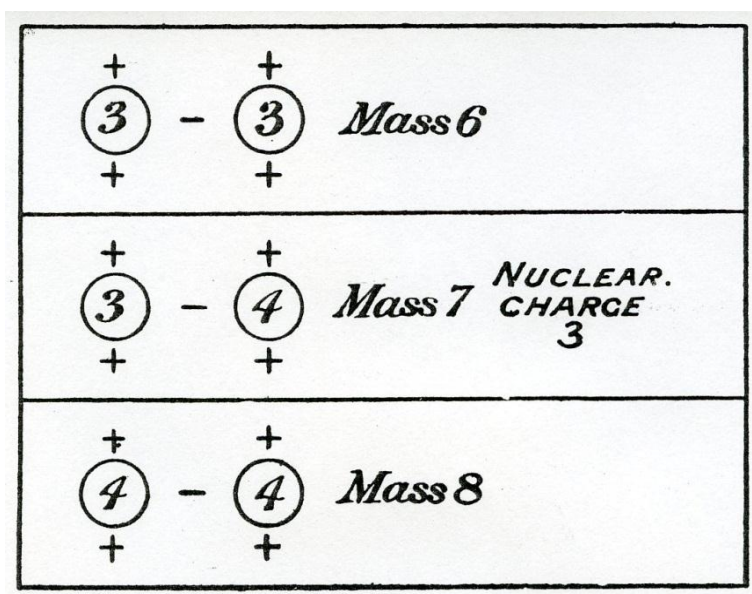

Figure 8. Rutherford's models for the three lithium isotopes, as he presented them in his 1920 Bakerian lecture. The negative signs represent nuclear electrons, and the circles with the numbers 1, 3, and 4 inside represent a proton, an $X_3^{++}$ particle, and an alpha particle,

At the same occasion he advocated the existence of the "neutron" as a tightly bound composite proton-electron particle. (The name "neutron" had been proposed as early as 1899, but Rutherford may have been unaware of it and he only used the name in 1921.) "Such an atom," he said, "would have very novel properties. … It should enter readily the structure of atoms, and may either unite with the nucleus or be disintegrated by its intense field, resulting possibly in the escape of a charged H atom or an electron or both" (Rutherford 1920, p. 34). And this was not all, for he also suggested that there was experimental evidence for a light helium nucleus consisting of three protons bound by one electron. We would denote such a particle $^3$He, but Rutherford used the notation $X_3^{++}$. Moreover, he saw no reason why a heavy hydrogen isotope ($^2$H) consisting of two protons and one electron – a deuteron in later terminology – should not exist (Figure 8). A couple of quotations from the 1920 Bakerian lecture may serve to illustrate Rutherford's ideas:

> In considering the possible constitution of the elements, it is natural to suppose that they are built up ultimately of hydrogen nuclei and electrons. … We have shown that atoms of mass about 3 carrying two positive charges are liberated by $\alpha$–particles both from nitrogen and oxygen, and it is natural to suppose that these atoms are independent units in the structure of both gases. … We have seen that so far the nuclei of three light atoms have been recognized experimentally as probable units of atomic structure, viz., $H_1^+$, $X_3^{++}$, $He_4^{++}$, where the subscript represents the mass of the element.

Rutherford's ideas of the structure of the nucleus were admittedly speculative and, he emphasized, "purely illustrative." He nonetheless took them seriously and continued for several years to develop them into an elaborate "satellite" model of the atomic nucleus. According to this model, the nucleus consisted of a massive core surrounded by proton and electron satellites as well as neutral satellites consisting of these two particles (Stuewer 1986). He believed that the satellite model explained not only the disintegration of light nuclei but also natural radioactivity, and found it important enough to include it in *Radiations from Radioactive substances*, a book published in 1930 and written with Chadwick and Charles Ellis. However, at that time it was realized that the satellite model was a blind alley and that no proper understanding of the internal structure of the atomic nucleus had been achieved by the large amount of work done through the 1920s.

Only in 1932, with Chadwick's discovery of the real neutron – an elementary particle and not a proton-electron composite – did physicists slowly begin to understand the nucleus that Rutherford had discovered more than twenty years earlier. And they realized that the nucleus, and the atom as a whole, could only be understood on the basis of the new quantum mechanics. Rutherford, on the other hand, largely ignored quantum theory in his extensive investigations of radioactivity and nuclear physics. It was not a kind of theory he appreciated.

## 7. Concluding remark

Rutherford continued to contribute to the frontier of nuclear physics until his death in 1937. Following the discovery in 1932 of deuterium, the mass-2 isotope of hydrogen, he engaged in a research programme with the aim of finding the suspected mass-3 isotope (tritium) or its nucleus, called the triton. From a historical point of view it is interesting that the possible existence of tritium was suggested as early as 1913, when Bohr made the suggestion at the Birmingham meeting of the British Association for the Advancement of Science (Kragh 2011; Kragh 2012, p. 97). He even tried to detect it by means of spectroscopy, but in vain.

About twenty years later Rutherford and his Cavendish group took up the question, which they did by bombarding deuterium atoms with accelerated deuterons. They hoped to detect the reaction

$$\,^2_1\mathrm{H} + \,^2_1\mathrm{H} \rightarrow \,^4_2\mathrm{He} \rightarrow \,^3_1\mathrm{H} + \,^1_1\mathrm{H}\,,$$

but were unable to confirm the presence of tritons. In his very last scientific paper, Rutherford (1937) carefully evaluated the evidence for and against tritium, concluding that although tritons had probably been detected in nuclear reactions, tritium had not been obtained in such quantities that its properties could be studied by ordinary physical and chemical methods. Two years later tritium was discovered in a cyclotron experiment, produced by means of the same deuteron-deuteron process and identified by its beta decay into helium-3 (Eidinoff 1948).